\documentclass[preprints,article,accept,pdftex,oneauthor,particles]{Definitions/mdpi} 

\firstpage{442} 
\pubvolume{5}
\issuenum{3}
\articlenumber{0}
\pubyear{2022}
\copyrightyear{2022}
\datereceived{6 October 2022} 
\dateaccepted{20 October 2022} 
\datepublished{22 October 2022} 
\hreflink{\scriptsize https://\linebreak doi.org/10.3390/particles5040034} 

\usepackage{color,bm}

\Title{Electromagnetic response in an expanding quark-gluon plasma}

\TitleCitation{Electromagnetic response in an expanding quark-gluon plasma}

\Author{Igor A. Shovkovy $^{1,2}$\orcidA{}}

\AuthorNames{Igor A. Shovkovy}

\AuthorCitation{Shovkovy, I. A.}

\address{%
$^{1}$ \quad College of Integrative Sciences and Arts, Arizona State University, Mesa, Arizona 85212, USA; igor.shovkovy@asu.edu
\\
$^{2}$ \quad Department of Physics, Arizona State University, Tempe, Arizona 85287, USA}

\abstract{The validity of conventional Ohm's law is tested in the context of a rapidly evolving quark-gluon plasma produced in heavy-ion collisions. Here we discuss the electromagnetic response using an analytical solution in kinetic theory. As conjectured previously, after switching on an electric field in a nonexpanding plasma, the time-dependent current is given by $\mathbf{J}(t)=(1-e^{-t/\tau_0}) \sigma_{0} \mathbf{E}$, where $\tau_0$ is the transport relaxation time and $\sigma_{0}$ is the steady-state electrical conductivity. Such an incomplete electromagnetic response reduces the efficiency of the magnetic flux trapping in the quark-gluon plasma and may prevent the observation of the chiral magnetic effect. Here we extend the study to the case of a rapidly expanding plasma. We find that the decreasing temperature and the increasing transport relaxation time have opposite effects on the electromagnetic response. While the former suppresses the time-dependent conductivity, the latter enhances it.}

\keyword{quark-gluon plasma; heavy-ion collisions; kinetic theory; transport; electrical conductivity} 
 
\begin{document}


\section{Introduction}

The relativistic heavy-ion collision experiments in Brookhaven and CERN produce the hottest state of matter ever created in experiments \cite{STAR:2005gfr,PHENIX:2004vcz}. It is so hot that not only nuclei but also their constituent protons and neutrons melt away. The corresponding state of matter is the quark-gluon plasma (QGP) \cite{Rischke:2003mt,Yagi:2005yb}. Over the last two decades, we learned much about its physical properties. The QGP produced by relativistic heavy-ion collisions has a rather high temperature of several hundred megaelectronvolts. While composed of deconfined quarks and gluons, it remains surprisingly strongly interacting. The strong interaction is responsible for a quick equilibration of the plasma, its high opacity to passing relativistic jets \cite{ATLAS:2010isq,CMS:2011iwn}, low viscosity \cite{Heinz:2013th}, and a well-pronounced hydrodynamic flow \cite{ALICE:2010suc}. Theoretical studies also predict that the QGP may reveal unusual features connected with the chiral magnetic and chiral separation effects \cite{Kharzeev:2013ffa,Liao:2014ava,Miransky:2015ava,Kharzeev:2015znc} that have roots in the quantum chiral anomaly \cite{Bell:1969ts,Adler:1969gk}.

The presence of background magnetic fields is one of the prerequisites for the chiral anomalous effects \cite{Kharzeev:2013ffa,Liao:2014ava,Miransky:2015ava,Kharzeev:2015znc}. Strong magnetic fields are indeed natural to expect in relativistic heavy-ion collisions. Since the colliding ions carry positive charges, they produce large electric currents while moving past each other at speeds close to the speed of light in opposite directions. According to theoretical estimates, the corresponding currents induce magnetic fields with the strengths of the order of $|eB|\sim m_\pi^2$ \cite{Skokov:2009qp,Voronyuk:2011jd,Bzdak:2011yy,Deng:2012pc}. 

The detailed description of QGP is intricate because the magnetic fields in heavy-ion collisions are so short-lived. Moreover, while the fields spike to large values at the moment of the closest approach of colliding ions, they may become negligible when the proper QGP forms and becomes equilibrated \cite{McLerran:2013hla}. If it is the case, the chiral effects would not have enough time to build up. Then, in turn, their observable signatures will be suppressed or non-existent.

One may suggest that the magnetic flux can be trapped and sustained by the QGP because the latter has a substantial electrical conductivity \cite{Tuchin:2013ie,Tuchin:2013apa,Gursoy:2014aka,Tuchin:2015oka}. If true, the field strength would decrease relatively slowly with time and, thus, remain sufficiently large to yield discernible effects due to the chiral anomaly. For a while, this scenario seemed plausible although, perhaps, only marginally so \cite{McLerran:2013hla,Yan:2021zjc}. On the other hand, it was even suggested that the electromagnetic response is too weak for applicability of the classical treatment \cite{Zakharov:2014dia}. A recent study in Ref.~\cite{Wang:2021oqq} subjected the underlying mechanism to another scrutiny. It claimed that the incomplete electromagnetic response in heavy-ion collisions, termed colloquially as the "violation of the conventional Ohm's law," would reduce trapping of the magnetic flux and, thus, strongly suppress possible observables due to the chiral anomalous effects. A similar conclusion follows also from other considerations in Ref.~\cite{Grayson:2022asf}. 

In retrospect, it should not be surprising that the conventional Ohm's law is violated on time scales shorter than the transport relaxation time. However, one needs to clarify why the electrical conductivity should be suppressed compared to the conventional Ohm's law in the steady state. Recall that the electrical conductivity quantifies the electromagnetic response to an electric field, which is a dissipative process. The dissipation comes from the momentum relaxation of charges scattering on one another. On times shorter than the relaxation time, the probability of scattering is negligible, no momentum relaxation occurs, and no dissipation is expected. In other words, the transport is effectively ballistic. Nevertheless, on short times scales, the electromagnetic response is indeed incomplete. Below we reconfirm the result by using analytical solutions within the framework of kinetic theory. We also extend the study to the case of QGP with a rapid expansion. 

The paper is organized as follows. In Sec.~\ref{sec:no-expansion}, we discuss the electromagnetic response in the simplest model of a plasma without expansion. The effects of expansion are studied in Sec.~\ref{sec:with-expansion}, where we address the effects of changing temperature, electric field, and collision rate. Discussion of the main results and the summary of findings are given in Sec.~\ref{sec:summary}

\section{Electromagnetic response in a plasma without expansion}
\label{sec:no-expansion}
 
Kinetic theory is a convenient tool for describing the electromagnetic response in a plasma. Within such a framework, a state of QGP is described by distribution functions for every particle type. For simplicity, we will restrict our consideration below to a single species of charged particles. Conceptually, such a model will capture the main qualitative features of the electromagnetic response. Also, the generalization to the case of multiple species is straightforward.

To study the electromagnetic response, we will perturb the plasma by applying a background electric field $\vec{E}$, which turns on suddenly at $t=0$.  Without loss of generality, we will assume that the electric field points in the $x$ direction. Such a field will induce a nonzero electric current in the plasma,
\begin{equation}
J_x(t)  = 4 e \int \frac{d^3 p}{(2\pi)^3} v_x f(p_x,t) ,
\label{current-definition}
\end{equation}
where $f(p_x,t)$ is a time-dependent distribution function in the perturbed plasma. By definition, the particle velocity is given by $v_x \equiv \partial \epsilon_p/\partial p_x $, where $\epsilon_p = \sqrt{p_x^2+p_\perp^2+m^2}$. Note that the extra overall factor 4 accounts for the contributions from both particles and antiparticles, as well as from the two spin degrees of freedom. 

The out-of-equilibrium distribution function $f(p_x,t)$ satisfies the following kinetic equation:
\begin{equation}
\frac{\partial f(p_x,t)}{\partial t} +eE_x \frac{\partial f(p_x,t)}{\partial p_x} =-\frac{f(p_x,t)-f_0(p_x)}{\tau_0},
\label{KE-no-expansion}
\end{equation}
where $E_x$ is the $x$-component of the electric field, $\tau_0$ is the transport relaxation time, and $f_0(p_x)= 1/(e^{\sqrt{p_x^2+p_\perp^2+m^2}/T_0}+1)$ is the Fermi-Dirac distribution function of the plasma in equilibrium at temperature $T_0$. It is appropriate to mention that, in heavy-ion collisions, the distribution function may differ considerably from the Fermi-Dirac one, especially during the early stages of plasma evolution. Thus, one might be tempted to complicate the analysis by using another, perhaps more realistic, pre-equilibrium distribution function. To gain a qualitative insight into the effects of rapid expansion, however, it suffices to use the Fermi-Dirac distribution.

In kinetic equation (\ref{KE-no-expansion}), we use the simplest relaxation-time approximation to model the collision integral. Largely such an approximation is justified in studies of electrical conductivity. Indeed, the corresponding transport is dominated by large-angle scattering processes, which can be characterized effectively by the transport relaxation time alone. For the validity of such an approximation in the context of QCD, we refer the reader to a representative microscopic study in Ref.~\cite{Greif:2014oia}, using the Boltzmann Approach to Multi-Parton Scatterings (BAMPS) model.

We will assume that the plasma was in equilibrium at $t=0$. Then, the initial condition for the distribution function reads $ f(p_x,0)=f_0(p_x)$. Taking it into account, we obtain the following time-dependent solution:
\begin{eqnarray}
f(p_x, t) &=& e^{-t/\tau_0}f_0(p_x- eE_x t)+\frac{1}{ \tau_0}\int_{0}^{t} e^{-t^\prime/\tau_0} f_0(p_x-eE_x t^\prime) dt^\prime \nonumber\\
&=&  f_0(p_x)-eE_x\int_{0}^{t} e^{-t^\prime/\tau_0} \frac{d f_0(p_x-eE_x t^\prime)}{dp_x} dt^\prime ,
\end{eqnarray}
where we integrated by parts to arrive the expression in the last line.

By substituting the solution into Eq.~(\ref{current-definition}), we derive the expression for the electric   current
\begin{eqnarray}
J_x(t)  &=& 4e \int \frac{d^3 p}{(2\pi)^3} \frac{p_x}{\sqrt{p_x^2+p_\perp^2+m^2}}  \left[ f_0(p_x)-eE_x\int_{0}^{t} e^{-t^\prime/\tau_0} \frac{d f_0(p_x-eE_x t^\prime)}{dp_x} dt^\prime  \right] \nonumber\\
&=& - 4e^2 E_x \int_{0}^{t} dt^\prime  e^{-t^\prime/\tau_0}\int \frac{d^3 p}{(2\pi)^3} \frac{p_x}{\sqrt{p_x^2+p_\perp^2+m^2}}   \frac{d f_0(p_x-eE_x t^\prime)}{dp_x} ,
\label{Jx-0}
\end{eqnarray}
where we took into account that the current vanishes in equilibrium. At the leading order, the expression is linear in the electrical field. Then, by comparing Eq.~(\ref{Jx-0}) with the standard transport relation $J_x(t)=\sigma(t) E_x$, where $\sigma(t)$ is the electrical conductivity, we extract the following time-dependent transport coefficient:
\begin{eqnarray}
\sigma(t)&\simeq & - 4 e^2 \int_{0}^{t} dt^\prime  e^{-t^\prime/\tau_0}\int \frac{d^3 p}{(2\pi)^3} \frac{p_x}{\sqrt{p_x^2+p_\perp^2+m^2}}   \frac{d f_0(p_x)}{dp_x} \nonumber\\
&=& (1-e^{-t/\tau_0})\sigma_{0} .
\label{sigma-0}
\end{eqnarray}
where $\sigma_{0}$ is the limiting value of conductivity at late times. It is nothing else but the conventional Ohm's conductivity. Without much limitation, one may assume that the particle mass is negligible ($m=0$). Then, for the Ohm's conductivity in the model at hand, we derive 
\begin{equation}
\sigma_{0} \simeq
-4 e^2 \tau_0  \int \frac{d^3 p}{(2\pi)^3} \frac{p_x}{\sqrt{p_x^2+p_\perp^2}}   \frac{d f_0(p_x)}{dp_x}  
=\frac{\tau_0}{9}  e^2 T_0^2. 
\label{sigma-Ohm}
\end{equation}
To make contact with QCD, let us compare this model result with more sophisticated calculations of electrical conductivity in Refs.~\cite{Greif:2014oia,Aarts:2014nba,Brandt:2015aqk,Ding:2016hua,Sahoo:2018dxn}. The corresponding results from different lattice calculations and partonic models are summarized in Ref.~\cite{Sahoo:2018dxn}. The compilation of such results suggests that the ratio $\sigma_{0} /T_{0}$ has a weak dependence on the temperature in the near-critical region. Therefore, by comparing with our model in Eq.~(\ref{sigma-Ohm}), we conclude that the transport relaxation time must be inversely proportional to the temperature, i.e., $\tau_0\propto T_0^{-1}$.

\begin{figure}[H]
\includegraphics[width=0.475\textwidth]{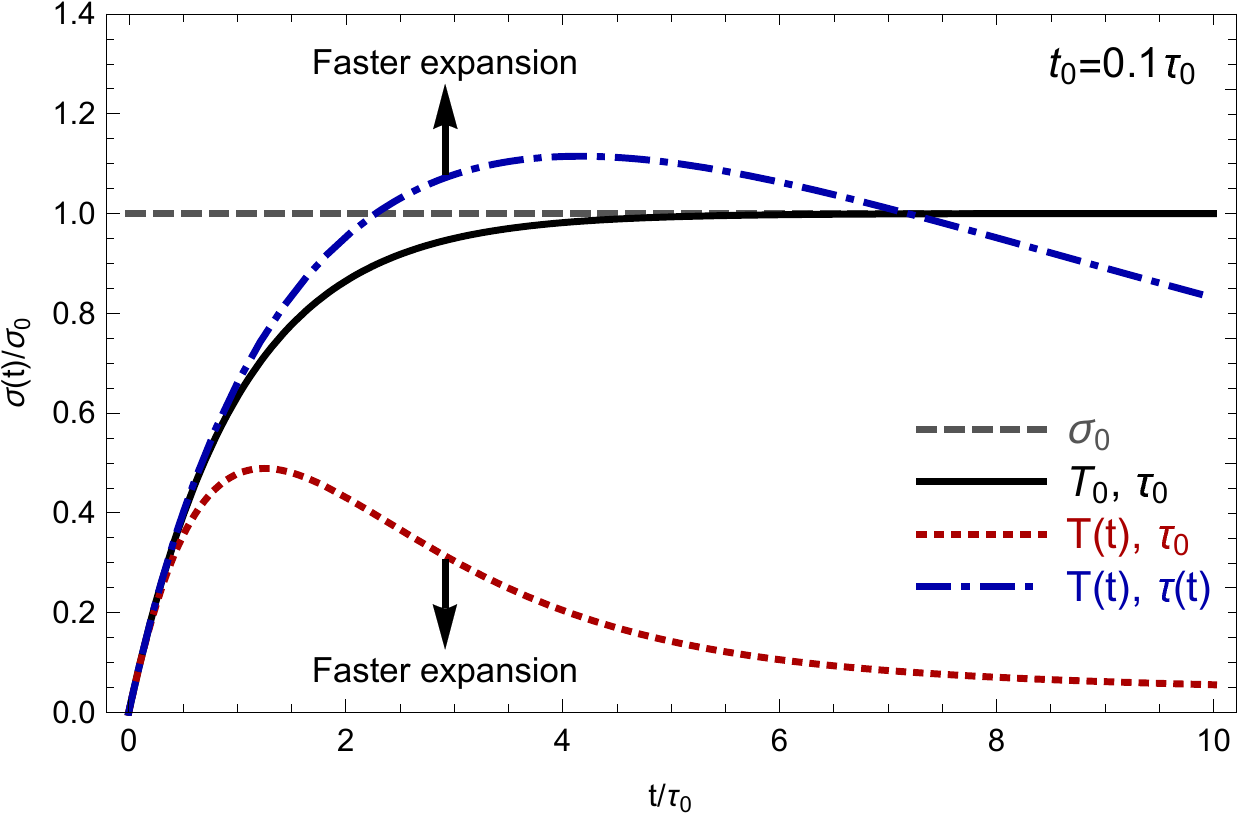}
\hspace{0.03\textwidth}
\includegraphics[width=0.475\textwidth]{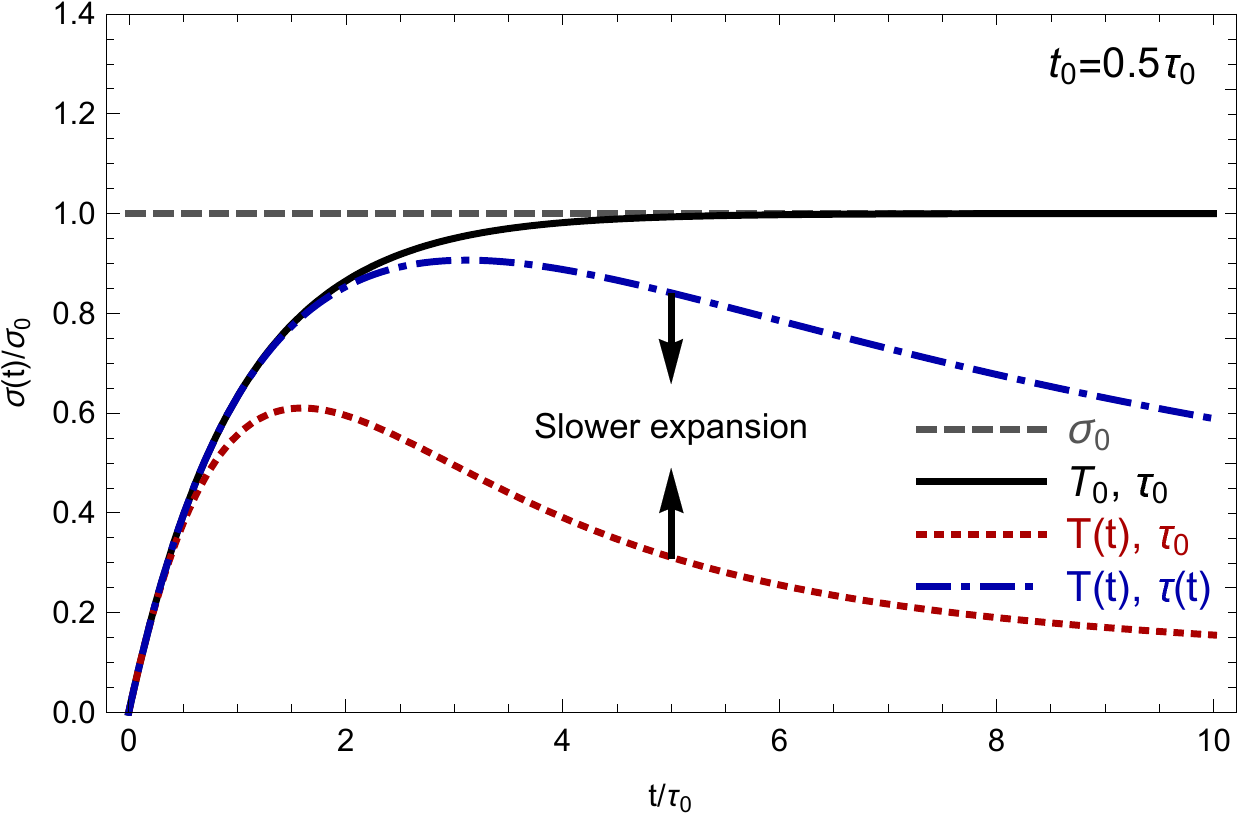}
\caption{The time-dependent conductivity in plasma without expansion (black solid line) and two models of expanding plasma (red dotted and blue dash-dotted lines). The dashed gray line shows the conventional Ohm's conductivity of a nonexpanding plasma. The expansion-driven time dependence of the temperature $T(t)$ and the transport relation time $\tau(t)$ is given in Eqs.~(\ref{T-vs-t}) and (\ref{tau-vs-t}), respectively.}
\label{sigma-time}
\end{figure}

The time-dependent response coefficient in Eq.~(\ref{sigma-0}) is compared with the late-time Ohm's conductivity $\sigma_{0}$ in Fig.~\ref{sigma-time}. As we see, there is a substantial suppression of the electromagnetic response (black solid line) on times up to about $t\sim 2\tau_0$. Such suppression is the realization of the so-called incomplete response in the plasma \cite{Wang:2021oqq,Grayson:2022asf}.

Within the kinetic theory, the underlying physics of the incomplete electromagnetic response is simple. It follows from the delayed response of charged particles to an external perturbation from the field. In other words, the electric current remains small when $t\lesssim \tau_0$ because the particles did not get enough time to accelerate to their "terminal" drift velocity. Clearly, the origin of such an incomplete response is purely kinematic and universal. One should also note that the momentum-relaxing scattering processes play no role on short time scales ($t\lesssim \tau_0$). 

\section{Electromagnetic response in a plasma with expansion}
\label{sec:with-expansion}

The description of the incomplete electromagnetic response in the previous section is instructive but hardly applicable directly to the QGP produced in heavy-ion collisions. In addition to its short lifetime, the plasma rapidly expands. Naturally, one may expect that the expansion drastically modifies the electromagnetic response.

Expansion of the plasma should have several physical consequences that may affect the electromagnetic response. The rapidly dropping temperature and the decreasing collision rate are the most obvious. Both can easily modify the incomplete response. Here we discuss how these effects change the plasma response within the kinetic theory.

\subsection{Effect of a decreasing temperature}

Instead of turning on both effects at once, it is instructive to consider how they add up one at a time. To start with, let us consider how a time-dependent temperature of an expanding plasma alone changes the electromagnetic response. 

As the plasma expands, its temperature decreases. Qualitatively, one may account for such an effect by replacing a constant temperature $T_0$ inside the equilibrium distribution function with a time-dependent one, i.e., 
\begin{equation}
T_t = T_0 \left(\frac{t_0+t}{t_0}\right)^{-\alpha}, \label{T-vs-t} 
\end{equation}
where $\alpha$ is a (positive) power. Its default value is $\alpha=1/3$ if one assumes the Bjorken expansion \cite{Bjorken:1982qr}.  The parameters $t_0$ and $T_0$ can be interpreted as the initial time and temperature of the QGP when an approximate local equilibration is reached. Qualitatively, a smaller (larger) value of $t_0$ corresponds to a faster (slower) expanding plasma. Note that we shifted the time coordinate so that the electric field is turned on at $t=0$ when the temperature of the locally equilibrated plasma is $T_0$. 

In this study, we concentrate primarily on the case of Bjorken expansion. It is sufficient for extracting the underlying qualitative physics affecting the electromagnetic response. Formally, the analysis also permits other non-Bjorken types of expansion when the value of parameter $\alpha$ is different from $1/3$.

In such a simplified model of an expanding plasma, the distribution function satisfies the following kinetic equation:
\begin{equation}
\frac{\partial f(p_x,t)}{\partial t} +eE_x \frac{\partial f(p_x,t)}{\partial p_x} =-\frac{f(p_x,t)-f_0(p_x,t)}{\tau_0},
\label{KE-expansion-1}
\end{equation}
where $f_0(p_x,t)= 1/(e^{\sqrt{p_x^2+p_\perp^2+m^2}/T_t}+1)$ is a quasi-equilibrium Fermi-Dirac distribution function with a time-dependent temperature $T_t$. By assumption, the transport relaxation time $\tau_0$ is time independent here.

The general solution of Eq.~(\ref{KE-expansion-1}) reads
 \begin{equation}
f(p_x, t) = e^{-t/\tau_{0}} \left[g\left(p_x- eE_x t\right)
+\frac{1}{\tau_{0}} \int_{0}^{t} e^{t^{\prime}/\tau_{0}} f_0\left(p_x- eE_x(t-t^{\prime}),t^{\prime}\right) dt^{\prime}\right],
\label{general-sol-1}
\end{equation}
where $g(p_x)$ is an arbitrary function. The latter is fixed by satisfying the initial condition for the distribution function, i.e., $ f(p_x,0)=f_0(p_x,0)$. It gives $g(p_x)= f_0(p_x,0)$. Then, by substituting this function into the general solution (\ref{general-sol-1}), we derive the following particular solution:
\begin{equation}
f(p_x, t) =e^{-t/\tau_{0}} \left[f_0\left(p_x- eE_x t,0\right)
+\frac{1}{\tau_{0}} \int_{0}^{t} e^{t^{\prime}/\tau_{0}} f_0\left(p_x- eE_x(t-t^{\prime}),t^{\prime}\right) dt^{\prime}\right] .
\end{equation}
Up to the linear order in the electric field, the approximate solution is given by
\begin{eqnarray}
f(p_x, t) &=&e^{-t/\tau_{0}} \left[f_0\left(p_x,0\right)
+\frac{1}{\tau_{0}} \int_{0}^{t} e^{t^{\prime}/\tau_{0}} f_0\left(p_x,t^{\prime}\right) dt^{\prime}\right] 
\nonumber\\
&& -eE_x e^{-t/\tau_{0}} \left[ t \frac{\partial f_0\left(p_x,0\right)}{\partial p_x}
+\frac{1}{\tau_{0}} \int_{0}^{t} e^{t^{\prime}/\tau_{0}} (t-t^{\prime}) \frac{\partial f_0\left(p_x,t^{\prime}\right) }{\partial p_x} dt^{\prime}
\right] .
\end{eqnarray}
By substituting the solution into Eq.~(\ref{current-definition}), we derive the leading order expression for the electric current in the form $J_x(t) =\sigma(t) E_x$, where 
\begin{equation}
\sigma(t) \simeq   -4e^2  \int \frac{d^3 p}{(2\pi)^3} \frac{p_x e^{-t/\tau_{0}} }{\sqrt{p_x^2+p_\perp^2+m^2 }}   \left[ t \frac{\partial f_0\left(p_x,0\right)}{\partial p_x}
+\frac{1}{\tau_{0}} \int_{0}^{t} e^{t^{\prime}/\tau_{0}} (t-t^{\prime}) \frac{\partial f_0\left(p_x,t^{\prime}\right) }{\partial p_x} 
dt^{\prime}
\right] .
\end{equation} 
After performing the integration over the momenta, we obtain 
\begin{eqnarray}
\sigma(t) &=& \frac{e^2T_0^2}{9} e^{-t/\tau_{0}} \left[t 
+ \frac{1}{\tau_{0}} \int_{0}^{t}e^{t^{\prime}/\tau_{0}} (t-t^{\prime}) \left(\frac{T_{t^{\prime}}}{T_0}\right)^2
dt^{\prime}
 \right] \nonumber\\
 &=& \frac{e^2T_0^2}{9} \int_{0}^{t}e^{(t^{\prime}-t)/\tau_{0}}  \left(\frac{T_{t^{\prime}}}{T_0}\right)^2\left(1+2\alpha \frac{t-t^{\prime}}{t_0+t^{\prime}} \right) dt^{\prime}.
\label{sigma-1b}
\end{eqnarray} 
In the last line, we integrated over $t^{\prime}$ by parts and took Eq.~(\ref{T-vs-t}) into account. In the special case of a time-independent temperature (i.e., $T_t = T_0$ and $\alpha =0$), the above conductivity expression reduces to the result in Eq.~(\ref{Jx-0}). However, it is very different when the temperature decreases due to expansion.

To estimate the quantitative effect of expansion on the electromagnetic response, we use two different values of the parameter $t_0$, namely $t_0=0.1 \tau_{0}$ ("fast" expansion) and $t_0=0.5 \tau_{0}$ ("slow" expansion). Both choices are reasonable for the QGP in heavy-ion collisions where $\tau_{0}$ is of the order of a few $\mbox{fm/c}$. The resulting time-dependent conductivity is shown with the red dotted line in Fig.~\ref{sigma-time}. As we see, the expansion-driven rapid decrease in the temperature leads to an additional strong suppression of the electromagnetic response.

\subsection{Effect of decreasing temperature and collision rate}

In the preceding subsection, we argued that the dropping temperature modifies the quasi-equilibrium state of an expanding plasma. Its consequence is a  strong suppression of the electromagnetic response. We ignored, however, the effect of the decreasing temperature on the collision rate. As is clear, the latter should go down along with the temperature. Since the transport relaxation time $\tau(t)$ is inversely proportional to the collision rate, function $\tau(t)$ must grow with time. Here we consider how it affects the response of the plasma.

By including the time dependence in the transport relaxation time, the kinetic equation for the distribution function becomes
\begin{equation}
\frac{\partial f(p_x,t)}{\partial t} +eE_x \frac{\partial f(p_x,t)}{\partial p_x} =-\frac{f(p_x,t)-f_0(p_x,t)}{\tau(t)},
\label{KE-expansion-2}
\end{equation}
where $f_0(p_x,t)$ is a quasi-equilibrium distribution function of the expanding plasma at time $t$ when its temperature is $T_t$. We model the time dependence of the temperature and the transport relaxation time as follows:
\begin{equation}
\tau =\tau_0 \left(\frac{t_0+t}{t_0}\right)^{\beta},\label{tau-vs-t} 
\end{equation}
where $\beta$ is a (positive) constant power. By recalling our arguments after Eq.~(\ref{sigma-Ohm}) that the transport relaxation time in QGP is approximately proportional to the inverse temperature, we will set $\beta=\alpha =1/3$. 

The general solution of Eq.~(\ref{KE-expansion-2}) reads
 \begin{equation}
f(p_x, t) = e^{-h(t)} \left[g\left(p_x- eE_x t\right)
+\int_{0}^{t} \frac{e^{h(t^{\prime})} }{\tau(t^{\prime})} f_0\left(p_x- eE_x(t-t^{\prime}),t^{\prime}\right) dt^{\prime}\right],
\label{general-sol-2}
\end{equation}
where $g(p_x)$ is an arbitrary function and 
\begin{eqnarray}
h(t) = \int_{0}^{t}\frac{d t^{\prime}}{\tau(t^{\prime})} 
=\frac{t_0}{\tau_0(1-\beta)}\left[\left(\frac{t_0+t}{t_0}\right)^{1-\beta}-1\right] .
\end{eqnarray}
By enforcing the initial condition $ f(p_x,0)=f_0(p_x,0)$, we derive $g(p_x)= f_0(p_x,0)$. Therefore, the particular solution reads
\begin{eqnarray} 
f(p_x, t) = e^{-h(t)} \left[f_0\left(p_x- eE_x t, 0\right)
+\int_{0}^{t} \frac{e^{h(t^{\prime})} }{\tau(t^{\prime})} f_0\left(p_x- eE_x(t-t^{\prime}),t^{\prime}\right) dt^{\prime}\right] .
\end{eqnarray}
To the linear order in $E_x$, the corresponding distribution function is given by
\begin{eqnarray} 
f(p_x, t)&\simeq & e^{-h(t)} \left[f_0\left(p_x, 0\right)
+\int_{0}^{t} \frac{e^{h(t^{\prime})} }{\tau(t^{\prime})} f_0\left(p_x,t^{\prime}\right) dt^{\prime}\right]
 \nonumber\\
&-& eE_xe^{-h(t)} \left[t \frac{\partial f_0 \left(p_x,0\right)}{\partial p_x} 
+\int_{0}^{t} \frac{e^{h(t^{\prime})} }{\tau(t^{\prime})} (t-t^{\prime}) \frac{\partial f_0 \left(p_x,t^{\prime}\right)}{\partial p_x} dt^{\prime}
 \right] .
\end{eqnarray}
Thus, by using of Eq.~(\ref{current-definition}), we derive the expression for the electric current in the form $J_x(t)  = \sigma(t)E_x $, where the time-dependent conductivity is 
\begin{equation}
\sigma(t) = -4e^2 \int \frac{d^3 p}{(2\pi)^3} \frac{p_x e^{-h(t)} }{\sqrt{p_x^2+p_\perp^2+m^2 }}  \left[t \frac{\partial f_0 \left(p_x,0\right)}{\partial p_x} 
+\int_{0}^{t} \frac{e^{h(t^{\prime})} }{\tau(t^{\prime})} (t-t^{\prime}) \frac{\partial f_0 \left(p_x,t^{\prime}\right)}{\partial p_x} dt^{\prime}
 \right].
 \end{equation}
After performing the integration over the momenta, we obtain 
\begin{eqnarray}
\sigma(t) &=& \frac{e^2T_0^2}{9} e^{-h(t)} \left[t 
+\int_{0}^{t} \frac{e^{h(t^{\prime})} }{\tau(t^{\prime})} (t-t^{\prime}) \left(\frac{T_{t^{\prime}}}{T_0}\right)^2
dt^{\prime}
 \right] \nonumber\\
&=& \frac{e^2T_0^2}{9} \int_{0}^{t}e^{h(t^{\prime})-h(t)}  \left(\frac{T_{t^{\prime}}}{T_0}\right)^2\left(1+2\alpha \frac{t-t^{\prime}}{t_0+t^{\prime}} \right) dt^{\prime} ,
 \label{sigma-2a}
 \end{eqnarray}
where, in the last line, we integrating over $t^{\prime}$ by parts and took Eq.~(\ref{T-vs-t}) into account.
 
The time-dependent conductivity (\ref{sigma-2a}) is shown with the blue dash-dotted line in Fig.~\ref{sigma-time}. The two panels dislpay the results for different sets of model parameters, i.e., $t_0=0.1 \tau_{0}$ ("fast" expansion) and $t_0=0.5 \tau_{0}$ ("slow" expansion). It is interesting, although perhaps not surprising, that the expansion-driven increase of the transport relaxation time enhances the electromagnetic response. One can also verify that the effect tends to grow with increasing the expansion rate.

\section{Discussion and Summary}
\label{sec:summary}

In this paper we studied the electromagnetic response in a rapidly evolving plasma, qualitatively similar to the one produced in heavy-ion collisions. Even in the absence of expansion, the response is incomplete on short time scales. It is manifested by the effective time-dependent conductivity, which is suppressed compared to the conventional transport coefficient in the Ohm law. This finding agrees with the previous studies \cite{Wang:2021oqq,Grayson:2022asf}. We argue that the corresponding regime has a purely kinematic origin. On the relevant short time scales, the transport is ballistic, with the momentum relaxation processes playing no much role. The current is suppressed, however, because it takes some time of order $\tau_0$ before the plasma particles accelerate to their "terminal" drift velocity.

Here we extended the analysis to account for a rapid expansion of the plasma, which is of interest in heavy-ion physics. As expected, the plasma expansion strongly modifies the electromagnetic response. We identified two major factors affecting the effective time-dependent electrical conductivity: (i) the decreasing temperature and (ii) the increasing transport collision time. To model the expansion, we used the Bjorken power-law scaling of temperature as a function of time. To include the effect of the rapid expansion on the transport relaxation time $\tau(t)$, we took into account the lattice estimates of the conductivity in the near-critical regime of QCD  \cite{Aarts:2014nba,Brandt:2015aqk,Ding:2016hua,Sahoo:2018dxn}. The latter suggests that the transport time $\tau(t)$ is approximately proportional to the inverse temperature. 

The decreasing temperature alone has an additional strong suppression effect on the electromagnetic response in an expanding plasma. On the other hand, the expansion-driven increase of the transport collision time has the opposite tendency of enhancing the electrical conductivity. We find that, in general, the interplay of the two effects can be nontrivial. If the expansion is moderately slow, the net result is an overall suppression of the response coefficient as compared to the case of a nonexpanding plasma. If the expansion is sufficiently fast, however, the enhancement from the increasing $\tau(t)$ can be substantial. In fact, at intermediate time scales, the time-dependent conductivity may even exceed the conventional Ohm value.

It might be instructive to emphasize that, despite using a simple relaxation-time approximation and assuming the Bjorken expansion, the qualitative insight into the underlying physics is very compelling. Indeed, irrespective of specific model details, not only does one expect a decreasing temperature but also an increasing transport collision time in an expanding plasma. The latter is the new effect that remained overlooked previously in studies of electrical conductivity.

In summary, the electromagnetic response of a short-lived and rapidly expanding plasma is very different from the conventional electrical conductivity. Overall the effective time-dependent conductivity is suppressed on sufficiently short time scales. It is a general feature shared by both nonexpanding and expanding plasmas. It implies that the decay of the magnetic field in the QGP plasma goes faster than the model with the conventional Ohm's conductivity suggests. Our study points to a possibility, however unlikely, that a sufficiently fast expansion can increase the transport time $\tau(t)$ enough to make the effective electrical conductivity at intermediate times larger than Ohm's value. In such a case, the decay rate of the magnetic field can be slowed down. Such a scenario will be invaluable for increasing the prospects of chiral magnetic effect in heavy-ion collisions. 

Our qualitative study offers a possibility of enhanced electrical conductivity in a rapidly expanding plasma. The physical consequences of such a prediction could be very interesting for heavy-ion physics. That said, it would be wise to accept the findings with caution. While the model used a trustworthy kinetic theory framework, the collision integral was implemented using the simplest relaxation-time approximation. In addition, the treatment of an expanding plasma was semi-rigorous. We utilized the simplest Bjorken scaling for the temperature and the transport collision time. We can hope only that future quantitative studies can verify our prediction of a possible enhancement of the electromagnetic response.


\funding{This research was funded by the U.S. National Science Foundation under Grant No.~PHY-2209470.}
 
\dataavailability{Not applicable.} 

\acknowledgments{The author thanks Kirill Tuchin for useful discussions.}

\conflictsofinterest{The author declares no conflict of interest.} 

\abbreviations{Abbreviations}{
The following abbreviations are used in this manuscript:\\

\noindent 
\begin{tabular}{@{}ll}
QGP & Quark-gluon plasma\\
QCD & quantum chromodynamics\\
\end{tabular}}

\appendixtitles{no} 

\begin{adjustwidth}{-\extralength}{0cm}

\reftitle{References}



\end{adjustwidth}
\end{document}